\magnification=1200
\baselineskip=18 pt
\font\bigbf=cmbx10 scaled 1400
\hfuzz 25pt

\centerline{\bigbf A Family of Gravitational Waveforms}
\centerline{\bigbf from Rapidly Rotating Nascent Neutron Stars
\footnote{$^1$}{\rm This note was written in 1996. It was not intended for
publication. Since I have been getting requests from
people interested in GW data analysis about the waveform
information, I thought it might be useful to put 
this note on gr-qc, so that I don't have to spend time looking for the TeX file
every time I get a request.}
}
\bigskip
\centerline{Dong Lai\footnote{$~^2$}{Current Address:
Space Sciences Bldg., Cornell University, Ithaca, NY 14853}}
\smallskip
\centerline{\it Theoretical Astrophysics, 130-33, California Institute
of Technology, Pasadena, CA 91125}
\smallskip
\centerline{(April 1996)}
\bigskip
\bigskip

This note describes fitting formulae for the gravitational waveforms
generated by a rapidly rotating neutron star (e.g., newly-formed 
in the core collapse of a supernova) as it evolves from 
an initial axisymmetric configuration toward a triaxial
ellipsoid (Maclaurin spheroid $\Rightarrow$ Dedekind ellipsoid).
This evolution is driven by the gravitational radiation 
reaction (a special case of the CFS instability). The details 
and numerical results can found in [Lai \& Shapiro, 1995, ApJ,
442, 259; Here referred as LS].

I will use the units such that $G=c=1$.

The waveform (including the polarization) is given by Eq.~(3.6)
of LS. Since the waveform is quasi-periodic, I will give fitting
formulae for the wave amplitude $h$ (Eq.~[3.7] of LS) 
and the quantity $(dN/d\ln f)$ (Eq.~[3.8] of LS; related to the
frequency sweeping rate), from 
which the waveform $h_+(t)$ and $h_\times(t)$ can be 
easily generated in a straightforward manner. 

\bigskip
{\bf Wave Amplitude:} The waveform is parametrized by three
numbers: $f_{max}$ is the maximum wave frequency in Hertz,
$M_{1.4}=M/(1.4M_\odot)$ is the NS mass in units of $1.4M_\odot$,
$R_{10}=R/(10\,{\rm km})$ is the NS radius in units of $10$ km. 
(Of course, the distance $D$ enters the expression
trivially.) It is convenient to express the dependence of $h$
on $t$ through $f$ (the wave frequency), with $f(t)$ to be determined 
later. A good fitting formula is 
$$h[f(t);f_{max},M,R]={M^2\over DR}A\left({f\over
f_{max}}\right)^{2.1}\left(1-{f\over f_{max}}\right)^{0.5},
\eqno(1a)$$
where 
$$A=\cases{(\bar f_{max}/1756)^{2.7}, & for $\bar f_{max}\le
500$~Hz;\cr
(\bar f_{max}/1525)^3, & for $\bar f_{max}\ge 330$~Hz;\cr}
~~~~~~{\rm with}~\bar f_{max}\equiv f_{max}R_{10}^{3/2}M_{1.4}^{-1/2}.
\eqno(1b)$$
Note that if we want real numbers, we have
$${M^2\over DR}=4.619\times 10^{-22}M_{1.4}^2R_{10}^{-1}
\left({30\,{\rm Mpc}\over D}\right).
$$

\bigskip
{\bf Number of wave cycles per logarithmic frequency:}
The fitting formula is 
$$\left |{dN\over d\ln f}\right|=\left({R\over M}\right)^{5/2}
{0.016^2 (R_{10}^{3/2}M_{1.4}^{-1/2} f/1\,{\rm Hz})\over
A^2(f/f_{max})^{4.2}[1-(f/f_{max})]}.
\eqno(2)$$

\bigskip
{\bf Note (i)}: using equations (1)-(2), we obtain the ``characteristic 
amplitude'':
$$\eqalign{h_c=h\left|{dN\over d\ln f}\right|^{1/2}
&=0.016{M^{3/4}R^{1/4}\over D}
\left({R_{10}^{3/2}M_{1.4}^{-1/2}\,f\over 1\,{\rm Hz}}\right)^{1/2}\cr
&=5.3\times 10^{-23}\left({30\,{\rm Mpc}\over D}\right)
M_{1.4}^{3/4}R_{10}^{1/4}\left({R_{10}^{3/2}M_{1.4}^{-1/2}
f\over 1\,{\rm Hz}}\right)^{1/2},
}$$
which agrees with Eq.~(3.12) of LS to within $10\%$ [Note that 
in Eq.~(3.12) of LS, the factor $f^{1/2}$ should be replaced by 
$(R_{10}^{3/2}M_{1.4}^{-1/2}\,f)^{1/2}$, similar to the above 
expression.]

\bigskip
{\bf Note (ii)}: The accuracy of these fitting formulae (as compared to 
the numerical results shown in LS) is typically within $10\%$. 
When $f$ is very close to $f_{max}$, the error in the fitting can be
as large as $30\%$.

\bigskip
{\bf Note (iii)}: The frequency evolution $f(t)$ is obtained 
by integrating the equation
$f^2/\dot f=-|dN/d\ln f|$ (note that the frequency sweeps from
$f_{max}$ to zero). For example, we can choose $t=0$ at
$f=0.9f_{max}$. (Note that one should not choose $t=0$ at
$f=f_{max}$ as the time would diverge --- the actual evolution 
near $f_{max}$ depends on the initial perturbations). 
	
	Once $f(t)$ is obtained, the waveform can be calculated as
(cf.~Eq.~[3.6] of LS):
$$\eqalign{
h_+ &=h[f(t);f_{max},M,R]\cos\Phi(t){1+\cos^2\theta\over 2},\cr
h_\times &=h[f(t);f_{max},M,R]\sin\Phi(t)\cos\theta,\cr
}$$
where $\theta$ is the angle between the rotation axis of the star 
and the line of sight from the earth, and 
$\Phi(t)=2\pi\int f(t)dt$ is the phase of the gravitational wave. 

\bigskip
{\bf Note (iv)}: $f_{max}$ typically ranges from $100$ Hz to $1000$ Hz
(see Fig.~5 of LS); $M_{1.4}$ and $R_{10}$ are of order unity for 
realistic neutron stars. 

\end